\documentclass[twocolumn,showpacs,preprintnumbers,amsmath,amssymb,superscriptaddress]{revtex4}
\usepackage{graphicx}
\usepackage{dcolumn}
\usepackage{bm}
\usepackage{soul}
\usepackage{color}
\usepackage{epstopdf}
\usepackage[version=3]{mhchem}
\usepackage{lipsum}
\usepackage[outercaption]{sidecap}
\usepackage{floatrow}
\begin{document}

\title{Casimir interactions in strained graphene systems}

\author{Anh D. Phan}
\affiliation{Department of Physics, University of Illinois, 1110 West Green St, Urbana, Illinois 61801, USA}
\email{anhphan2@illinois.edu}
\affiliation{Institute of Physics, 10 Daotan, Badinh, Hanoi, Vietnam}%
\author{The-Long Phan}
\affiliation{Department of Physics, Chungbuk National University, Cheongju 361-763, Korea}
\email{ptlong2512@yahoo.com}
\date{\today}

\begin{abstract}
We theoretically study the strain effect on the Casimir interactions in graphene based systems. We found that the interactions between two strained graphene sheets are strongly dependent on the direction of stretching. The influence of the strain on the dispersion interactions is still strong in the presence of dielectric substrates but is relatively weak when the substrate is metallic. Our studies would suggest new ways to design next generation devices.

\end{abstract}

\maketitle
The possiblity of strain engineering to tailor the optical and electrical properties of graphene has been of great interest to scientists. Graphene is currently one of the most fascinating materials, possessing notable features as well as much utilitity in nanodevices \cite{16,17}. The single-layer material's properties are highly sensitive to external fields, induced doping and applied stress. The band gap of graphene was shown to be dramatically modified under mechanical deformation \cite{1,2,3}. The ultratensible strain, up to 20 $\%$ can be applied on graphene without losing reversible elastic deformation \cite{4}. The strain dependence of the optical conductivity of graphene can be exploited to design graphene-based sensors \cite{5}.

The Casimir interaction plays an important role in fabricating and operating nano- and micro-electromechanical systems. This interaction derives from the electromagnetic fluctuations between objects. The attractive force due to the Casimir effect induces adhesion, stiction and friction in nanodevices \cite{13,19} and becomes significant at short distances. Finding approaches to control the magnitude of the Casimir force is of increasingly growing interest in the search to reduce and avoid such unwanted problems. In particular, studying the dispersion force in graphene-based systems not only provides fundamental understanding for nanoscience, but also opens up novel graphene-based applications.

There are numerous studies in the field of the Casimir interaction in graphene systems. Previous researchers have shown the possibility of obtaining the repulsive Casimir force when graphene systems are immersed in bromobenzene \cite{13} or graphene interacts with metamterials \cite{20}. Other papers show the possibility of changing the Casimir effect by means of applying an electric field to doped unstrained graphene sheet \cite{19,12,18}. 

In this letter, we have calculated, for the first time, the Casimir energy in strained graphene systems with different directions of stretching. Our calculations were carried out using the Lifshitz formula. We study how the Casimir interaction is influenced by the direction of strain as well as the strain modulus. In addition, we also show how the strain engineering on graphene affects the Casimir interactions between semi-infinite substrates.

The Casimir energy per unit area between two parallel flat bodies at  $T = 0$ $K$, separated by a distance $a$, is given by the Lifshitz formula \cite{10,11}
\begin{eqnarray}
E(a) =\frac{\hbar}{4\pi^2}\int_0^{\infty}qdq \int_0^{\infty}d\xi \ln det\left[1-\bm{R}_1\bm{R}_2e^{-2qa} \right],\nonumber\\
\end{eqnarray}
where $\hbar$ is the Planck constant, $q = \sqrt{k_{||}^2+\frac{\xi^2}{c^2}}$ is the wave vector perpendicular to the object surface, $k_{||}$ is the wave vector parallel to the surface, $c$ is the speed of light and $\xi$ is an imaginary frequency with $\omega = i\xi$. $\bm{R}_1$ and $\bm{R}_2$ are the reflection coefficient matrices of object 1 and 2, respectively, given by \cite{21}
\begin{eqnarray}
\bm{R}&=&\frac{1}{\Delta}\left( \begin{array}{rcl}
\tilde{R}_{ss} & \tilde{R}_{sp} \\ 
\tilde{R}_{ps} & \tilde{R}_{pp} \\
\end{array}\right),\nonumber\\
\Delta &=& \left(\frac{\varepsilon q}{k}+\frac{\sigma_{xx}q}{\varepsilon_0\xi} + 1\right)\left(1+\frac{k}{ q} +\frac{\mu_0\sigma_{yy}\xi}{q}\right)+\frac{\mu_0}{\varepsilon_0}\sigma_{xy}^2, \nonumber\\
\tilde{R}_{ss} &=& \left(\frac{\varepsilon q}{k}+\frac{\sigma_{xx}q}{\varepsilon_0\xi} +1\right)\left(1-\frac{k}{ q} -\frac{\mu_0\sigma_{yy}\xi}{q}\right)-\frac{\mu_0}{\varepsilon_0}\sigma_{xy}^2, \nonumber\\
\tilde{R}_{pp} &=& \left(\frac{\varepsilon q}{k}+\frac{\sigma_{xx}q}{\varepsilon_0\xi} -1\right)\left(1+\frac{k}{ q} +\frac{\mu_0\sigma_{yy}\xi}{q}\right)+\frac{\mu_0}{\varepsilon_0}\sigma_{xy}^2, \nonumber\\
\tilde{R}_{sp} &=& \tilde{R}_{ps} = 2\sqrt{\frac{\mu_0}{\varepsilon_0}}\sigma_{xy}, \\
\nonumber
\label{eq:1}
\end{eqnarray}
where $\mu_0$ is the permeability of free space, $\varepsilon_0$ is the vacuum permittivity, $k=\sqrt{k_{||}^2+\varepsilon\frac{\xi^2}{c^2}}$, $\varepsilon \equiv \varepsilon(i\xi)$ is the dielectric function of substrate as a function of imaginary frequencies, the subscripts $p$ and $s$ denote for transverse magnetic (TM) and transverse electric (TE) mode, respectively. $\sigma_{xx}$, $\sigma_{yy}$, $\sigma_{xy}$ and $\sigma_{yx}$ are graphene optical conductivities for each respective direction.

\begin{figure}[htp]
\includegraphics[width=6cm]{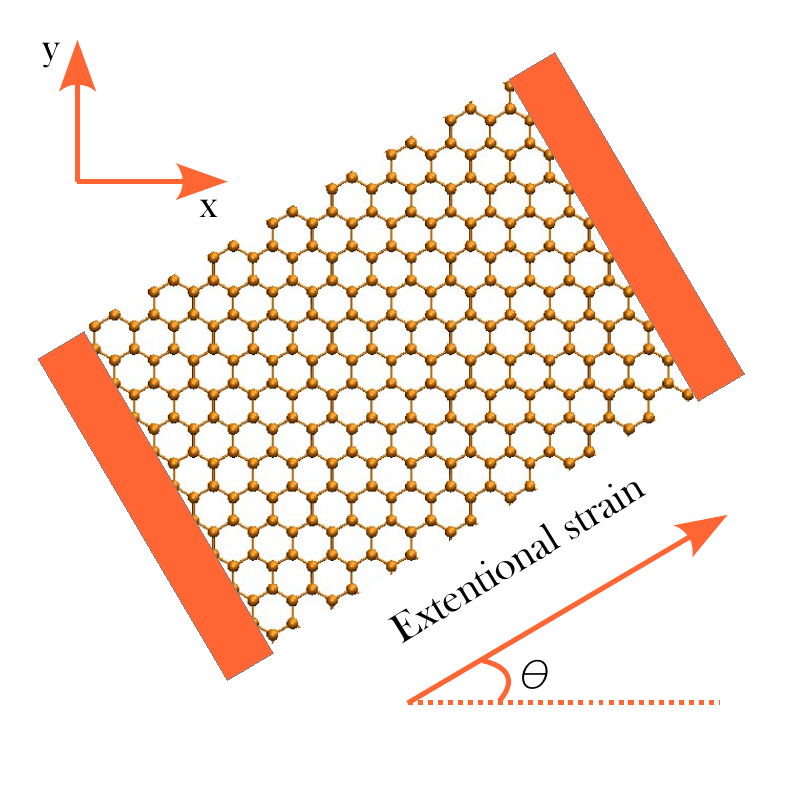}
\caption{\label{fig:0}(Color online) Illustration of strained graphene with angle $\theta$ made by the direction of extensional strain and the $x$-axis.}
\end{figure}

For unstrained and free-standing graphene, $\sigma_{xx}=\sigma_{yy} = \sigma(i\xi)$ and $\sigma_{xy}=\sigma_{yx} = 0$. The optical conductivity for low frequency ($\xi \le$ 3 eV) is well described by the Kubo formalism \cite{6}. This theory is consistent with experimental results at low temperature, and $\sigma(i\xi)$ becomes the universal conductivity $\sigma_0=e^2/4\hbar$.

Under the uniaxial strain, the optical conductivity of graphene can be expressed by \cite{7} 
\begin{eqnarray}
\sigma_{xx} = (1-2\gamma\epsilon_{xx})J\sigma_0 &,& \sigma_{yy} = (1-2\gamma\epsilon_{yy})J\sigma_0, \nonumber\\
\sigma_{xy} = &\sigma_{yx}& = -2\gamma\epsilon_{xy}J\sigma_0,
\label{eq:2}
\end{eqnarray}
where $\gamma = \beta - 1$ and $\beta = 3$ is the Gr$\ddot{u}$neisen parameter, $J$ is the Jacobian determinant calculated by $1/det(\bm{I}-\gamma\bm{\epsilon})$, and $\bm{I}$ is the identity matrix $2\times 2$. $\epsilon_{xx}$, $\epsilon_{yy}$ and $\epsilon_{xy}$ are strain components in the strain tensor $\bm{\epsilon}$ \cite{8}

\begin{eqnarray}
\bm{\epsilon}&=&\left( \begin{array}{rcl}
\epsilon_{xx} & \epsilon_{xy} \\ 
\epsilon_{yx} & \epsilon_{yy} \\
\end{array}\right)\nonumber\\
&=&\epsilon\left( \begin{array}{rcl}
\cos^2\theta - \nu\sin^2\theta & (1+\nu)\sin\theta\cos\theta \\ 
(1+\nu)\sin\theta\cos\theta & \sin^2\theta - \nu\cos^2\theta \\
\end{array}\right),
\label{eq:3}
\end{eqnarray}
where $\epsilon$ is the strain modulus, $\nu$ is the Poisson's ratio and $\theta$ is an angle made by the direction of the principal strain and the $x$-axis illustrated in Fig.\ref{fig:0}. The $x$-axis is defined as the axis parallel to the zigzag direction. For a graphene sheet, $\nu = 0.14$ is a value calculated by \emph{ab initio} simulations \cite{9}.

\begin{figure}[htp]
\includegraphics[width=8cm]{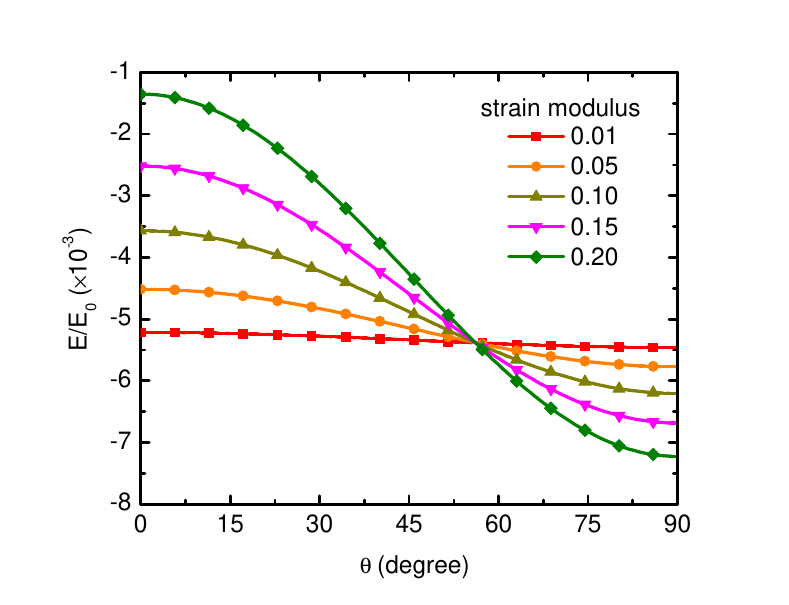}
\caption{\label{fig:1}(Color online) Relative Casimir energy between two graphene sheets normalized by the perfect metal $E_0=\pi^2\hbar c/720a^3$ as a function of $\theta$ with various values of the strain modulus.}
\end{figure}

To study the effect of uniaxial strains on the dispersion interactions, we calculate the Casimir energy between two graphene sheets versus the direction of stretching with respect to the $x$-axis at $a = 30$ nm and show results in Fig.\ref{fig:1}. The TM mode is known to have a much more significant role in the Casimir interaction at small distances compared to the TE mode. $\sigma_{xx}$ and $\sigma_{yy}$ are decisive factors for the TM mode and the TE mode, respectively. We found that at $\theta \approx 56^o$, $\sigma_{xx} = \sigma_0$ and the ratio of $E/E_0$ is independent of the strain modulus $\epsilon$. At the same angle with $\theta \le 56^o$, an increase of $\epsilon$ causes a decrease of $\sigma_{xx} < \sigma_0$. As a result, increasing $\epsilon$ leads to the reduction of the magnitude of the Casimir energy. For $\theta \ge 56^o$,  larger $\epsilon$ values result in greater values of $|E/E_0|$ because $\sigma_{xx} > \sigma_0$ is also larger.

For relatively small strain moduli ($\epsilon \le 0.01$), the ratio $E/E_0$ remains almost constant as the angle $\theta$ increases. This suggests that the strain effect weakly influences the Casimir energy between two graphene sheets. Applying more than 1 $\%$ mechanical strain allows us to significantly tune the Casimir force by rotating the direction of stretching. This finding also shows that the Casimir interactions are proportional to $1/a^3$, the same as the Casimir energy between two perfectly conducting plates. However, the calculated value $E/E_0 \approx -0.0053$ indicates that the dispersion energy in the two-graphene-sheet system is much smaller than that in the two-ideal-metal system, which is consistent with the previous study \cite{6}. 

To investigate the impact of the substrate on the Casimir energy as well as the strain effect of graphene on Casimir interaction between semi-infinite substrates, we consider the substrates made of silica and gold. For silica substrate, the dielectric function is described by the oscillator model \cite{13,14}
\begin{eqnarray}
\varepsilon (i\xi ) = 1 + \sum_{i}{\frac{{C _{i} }}{{1 + \xi^2/\omega_{i}^2 }}},
\end{eqnarray}
where $C_i$ and $\omega_i$ are an oscillator's strength and resonant frequency, respectively, in the $i$th mode. The parameter values are shown in Ref.\cite{14}. The parameter set and model show a good agreement between experimental data and theoretical calculations of the Casimir force between a silica plate and gold sphere immersed in bromobenzene \cite{14}.

For gold substrate, we use the Drude model to describe its dielectric function \cite{14,15}
\begin{eqnarray}
\varepsilon (i\xi ) = 1 + \frac{{\omega _p^2 }}{{\xi (\xi+\gamma) }},
\end{eqnarray}
where $\omega _p = 9.0$ eV, $\gamma = 0.035$ eV are the plasma frequency and the damping parameter of Au, respectively.

\begin{figure}[htp]
\includegraphics[width=8cm]{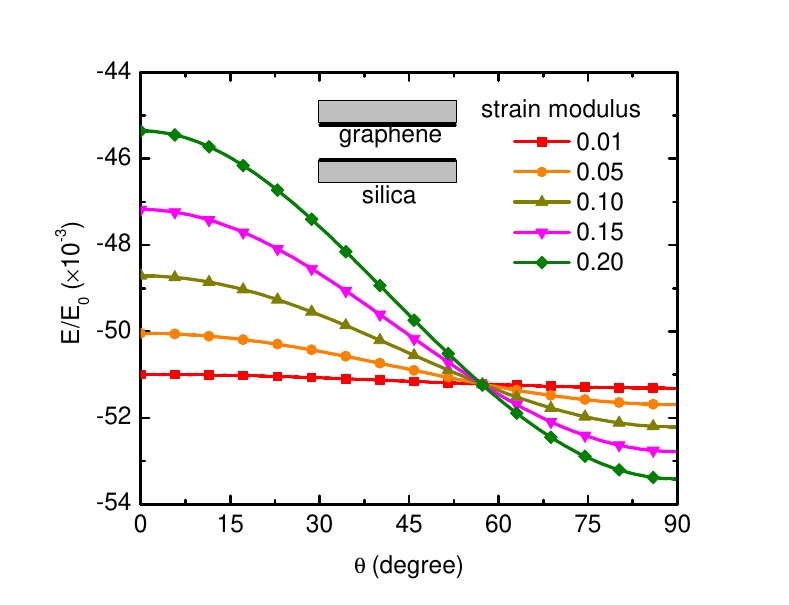}
\caption{\label{fig:2}(Color online) Relative Casimir energy between two strained graphenes on top of silica substrate normalized by the perfect metal $E_0$ as a function of $\theta$ with various $\epsilon$ values.}
\end{figure}

Figure \ref{fig:2} shows the Casimir energy between strained graphene sheets located on ${SiO_2}$ substrates at $a = 30$ nm. It was found that graphene has a significant influence on the Casimir interaction in such systems when the substrate is a dielectric material \cite{13,12}. Although silica substrates enhance this interaction, the features of Fig.\ref{fig:2} are similar to those of Fig.\ref{fig:1}. We see again that the Casimir energy has the same value for various tensible strains at $\theta \approx 56^o$. 


\begin{figure}[htp]
\includegraphics[width=8cm]{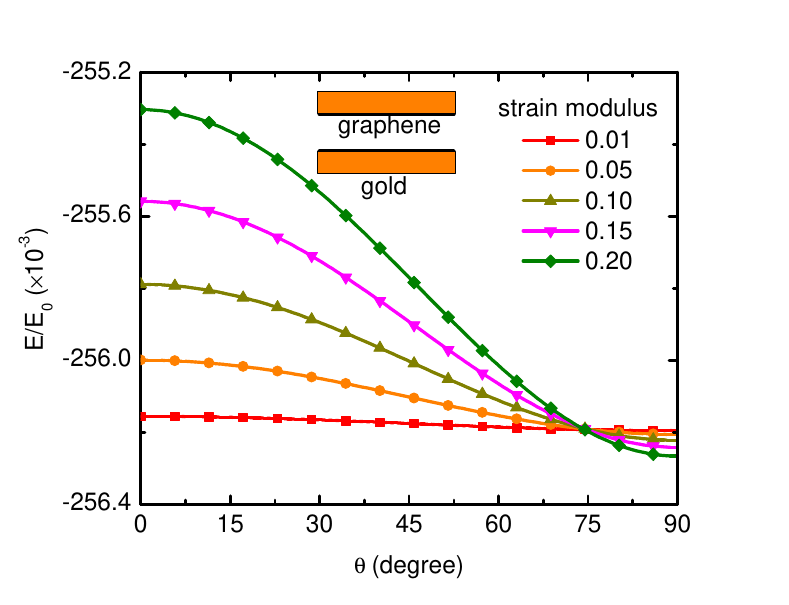}
\caption{\label{fig:3}(Color online) Relative Casimir energy between two strained graphene sheets on top of gold substrates, normalized by the perfect metal $E_0$ as a function of $\theta$ with various $\epsilon$ values.}
\end{figure}

The screening effect of a graphene coating on a metallic substrate has a weak influence on the Casimir interactions between two metal plates. As shown in Fig. \ref{fig:3}, the intersection point of the curves has shifted to $\theta$ $\approx$ $72^o$. However, the ratio $E/E_0$ diminishes slightly as $\theta$ increases, even at large $\epsilon$. $|(E_{max}-E_{min})/E_{max}| = 0.4$ $\%$ for the system of the strained graphene sheets on gold substrates. The value is much smaller than that in the case of the strained graphene sheets on silica substrates ($15.33$ $\%$). The small ratio $E/E_0$ is attributed to the transparency of graphene sheets when they are on gold substrates. This finding is consistent with a recent study \cite{12}. Authors in Ref.\cite{12} showed that an unstrained- and pristine-graphene coating has a small effect on metallic substrates. As a result, strain engineering on graphene cannot be exploited to tune the Casimir interaction in metal systems.

In summary, the Casimir interaction in the strained graphene systems has been theoretically investigated using the Lifshitz theory. Our model shows that the dispersion interaction heavily depends on the graphene optical conductivities significantly varied by applying the mechanical strain. Changing the strain modulus or the direction of applied strain with respect to the zigzag direction considerably modifies the Casimir interaction between two strained graphene sheets with and without silica substrate. However, at $\theta \approx 56^o$ for graphene sheets with and without silica substrates, and $\theta \approx 72^o$ for graphene sheets on gold substrates, the Casimir energy is not influenced by the stretching. As gold substrates coated by strained graphene sheet, the ratio $E/E_0$ nearly remains constant. This indicates that there is no way to tailor the Casimir interaction in metal systems using graphene. Our findings are extremely useful for the exploration of state-of-the-art applications.

\end{document}